\documentclass[preprint,onecolumn,superscriptaddress,nofootinbib]{revtex4}

\usepackage{rotating}
\usepackage{amsmath,amstext,amsbsy,amscd,bbm,epsfig,lscape}
\usepackage{graphicx,amsfonts,amssymb,color,comment,float,times}

\usepackage[utf8]{inputenc}

\definecolor{MyGrey}{rgb}{0,0,0} %defining the color 'MyDarkGreen'
\definecolor{MyDarkBlue}{rgb}{0.,0.,1} %defining the color 'MyDarkBlue'
\definecolor{MyLightBlue}{rgb}{0.22,0.51,0.9}

\usepackage[colorlinks=true,linkcolor=MyDarkBlue,citecolor=MyDarkBlue,urlcolor=MyLightBlue,bookmarksnumbered=true,bookmarksopen]{hyperref}

\usepackage{mathrsfs,amssymb,slashed}  
\usepackage{cancel}
\usepackage[normalem]{ulem}

\newcommand{\be}{\begin{equation}}
\newcommand{\ee}{\end{equation}}
\newcommand{\bea}{\begin{eqnarray}}
\newcommand{\eea}{\end{eqnarray}}

\begin{document}
%%%%%%%%%%%%%%%%%%%%%%%%%%%%%%%%%%%%%%%%%%%%%%%%%%%%%%%%%%%  FRONT PAGE

\title{Heavy Neutrino searches through Double-Bang Events at Super-Kamiokande, DUNE, and Hyper-Kamiokande }

\author{M.C. Atkinson}
\affiliation{TRIUMF, Vancouver, BC V6T 2A3, Canada}
\email{matkinson@triumf.ca}

\author{Pilar Coloma}
\affiliation{Instituto de Fisica Teorica UAM/CSIC, Calle Nicolas Cabrera 13-15, Universidad Autonoma de Madrid, 28049 Madrid, Spain}
\email{pilar.coloma@ift.csic.es}

\author{Ivan Martinez-Soler}
\email{ivan.martinezsoler@northwestern.edu}
\affiliation{Fermi National Accelerator Laboratory, Batavia, IL 60510, USA}
\affiliation{Northwestern University, Evanston, IL 60208, USA}
\affiliation{Colegio de F\'isica Fundamental e Interdisciplinaria de las Am\'ericas (COFI), 254 Norzagaray street, San Juan, Puerto Rico 00901.}

\author{Noemi Rocco}
\affiliation{Fermi National Accelerator Laboratory, Batavia, IL 60510, USA}
\email{nrocco@fnal.gov}
\author{Ian M. Shoemaker}
\email{shoemaker@vt.edu}
\affiliation{Center for Neutrino Physics, Department of Physics, Virginia Tech University, Blacksburg, VA 24601, USA}

\date{\today}
\begin{abstract}
A variety of new physics scenarios allows for neutrinos to up-scatter into a heavy neutral lepton state. For a range of couplings and neutrino energies, the heavy neutrino may travel some distance before decaying to visible final states. When both the up-scattering and decay occur within the detector volume, these  ``double bang'' events produce distinctive phenomenology with very low background. In this work, we first consider the current sensitivity at Super-Kamiokande via the atmospheric neutrino flux, and find current data may already provide new constraints. We then examine projected future sensitivity at DUNE and Hyper-Kamiokande, including both atmospheric and beam flux contributions to double-bang signals. 
\end{abstract}

\preprint{IFT-UAM/CSIC-21-59, FERMILAB-PUB-21-239-T, NUHEP-TH/21-04}

%\keywords{}

%%%%%%%%%%%%%%%%%%%%%%%%%%%%%%%%%%%%%%%%%%%%%%%%%%%%%%%%%%%%%%%%%%%
\maketitle

\section{Introduction.}

The nature of neutrino masses remains mysterious and requires physics beyond the Standard Model (SM) of particle physics. One of the simplest scenarios leading to the generation of light neutrino masses invokes right-handed sterile neutrinos (also known as heavy neutral leptons), $N$, which are singlet states under the SM gauge interactions. In this case, the neutrino mass Lagrangian reads
\be
\mathcal{L}_{mass}^{\nu} \supset Y_{\nu} \overline{L}_L \tilde\phi N_R + \frac{1}{2}M_R \overline{N_R^c}N_R \, + \textrm{h.c.} \, ,
\label{eq:seesaw}
\ee
where $\tilde\phi \equiv i\sigma_2 \phi^*$, $ N_R^c \equiv C \bar N_R^T$ is the charge conjugate of $N_R$ and we have omitted flavor and mass indices. This is the well-known Type I Seesaw Lagrangian\cite{Minkowski:1977sc,GellMann:1980vs,Mohapatra:1979ia}. Upon electroweak symmetry breaking, the left- and right-handed neutrinos mix, and the sterile neutrinos effectively inherit a reduced coupling to the weak force through the mixing matrix parameters. Therefore, in this scenario a HNL could be obtained using the same production mechanisms as for the SM neutrinos, provided that its mass is below that of the parent meson. The mixing mediates both the production as well as the decay of the heavy neutrino back into SM particles and, being weakly coupled, the HNL will be typically long-lived. 

It should be kept in mind that although the Lagrangian in Eq.~\eqref{eq:seesaw} implies a Majorana HNL, the conclusions from this study hold regardless of its Dirac/Majorana nature. The mass of the HNL is obtained after diagonalization of the mass Lagrangian and will eventually depend on the values of the Yukawa couplings and the Majorana mass in Eq.~\eqref{eq:seesaw}. Here, for simplicity, we follow a phenomenological approach and denote as $U$ the leptonic mixing matrix, and $m_N$ the mass of the HNL. As we will see, mixing at the level $|U|^{2} \sim 10^{-2} - 10^{-3}$, lead to HNL decay lengths around the size of the detector. 

For HNL masses in the GeV range, fixed-target experiments set the best constraints on the mixing of the HNL with electron and muon neutrinos, see e.g. Refs.~\cite{Atre:2009rg,Bryman:2019ssi,Bryman:2019bjg,Drewes:2015iva}. Moreover the DUNE near detector complex should be able to significantly improve on these~\cite{Ballett:2019bgd,Berryman:2019dme,Coloma:2020lgy,Breitbach:2021gvv}. However, if the HNL mixes primarily with tau neutrinos then its existence is much more difficult to probe using this approach: the HNL would have to be produced either in $\tau$ decays or in the decays of heavy mesons with a sizable branching ratio into $\nu_\tau$ (such as $D$ or $D_s$), which have a much smaller production cross section than lighter mesons. For $m_N \sim \mathcal{O}(0.1-10)~\textrm{GeV}$ the strongest constraints come from the DELPHI~\cite{Abreu:1996pa} and CHARM~\cite{Orloff:2002de} experiments, but values of $|U_{\tau 4}|^2 \sim 10^{-2}$ are still allowed for a heavy neutrino with mass around $m_N \sim \mathcal{O}(400)~\textrm{MeV}$~\cite{Atre:2009rg}. Note that light HNLs below the CHARM bound, can be probed via terrestrial up-scattering followed by HNL decay in large volume detectors~\cite{Plestid:2020ssy}. 

However, it is important to stress that this is not the only possibility for a HNL to interact with the visible sector. For example, the active and HNL states 
may be coupled via a transition dipole moment. At low energies this is described by an effective operator
\be
\label{eq:dipole}
\mathcal{L}^\nu_{{\rm dipole}} \supset \mu ( \bar{\nu}_{L} \sigma^{\mu\nu} N) F_{\mu \nu}+ \textrm{h.c.} \, ,
\ee
where $\mu$ is a coefficient with units $({\rm mass})^{-1}$ which controls the strength of the interaction, while $F_{\mu\nu}$ is the electromagnetic field strength tensor, and $\sigma^{\mu\nu} \equiv \frac{i}{2}\left[ \gamma^\mu, \gamma^\nu\right]$. Although, in this scenario, a distinction between the Dirac or Majorana nature of the neutrinos is possible by looking at the angular distribution of the heavy neutrino decay~\cite{Li:1981um, Balantekin:2018ukw, Shrock:1982sc}, we will not address this question in this work.

Compared with their electron and muon counterparts,  tau neutrinos are more challenging to produce since they come from either $\tau$ or heavy meson decays, as outlined above. Consequently, the resulting bounds on transition magnetic moments which dominantly couple the HNL to the $\nu_{\tau}$ are significantly weaker than those coupling to either $\nu_{e}$ or $\nu_{\mu}$~\cite{Gninenko:2009ks,Coloma:2017ppo,Magill:2018jla,Shoemaker:2018vii,Shoemaker:2020kji,Brdar:2020quo,Plestid:2020vqf,Vergani:2021tgc}. 

Previously, in Ref.~\cite{Coloma:2017ppo} it was shown that both such models can be probed using double-bang (DB) events at IceCube with atmospheric neutrinos. These DB events occur when an incoming neutrino $\nu$ scatters on a target $T$ in the detector: 
\bea
\nu + T \longrightarrow {\rm cascade} ~+~ &N& \\
&\hookrightarrow &N \rightarrow {\rm cascade}
\eea

In this paper, we extend the work of Ref.~\cite{Coloma:2017ppo} by computing the DB sensitivity at Super-Kamiokande (SK), DUNE, and Hyper-Kamiokande (HK). Despite their smaller detector volume when compared to Icecube/DeepCore, they offer several advantages:
\begin{description}
\item[Sensitivity to smaller separation between the two bangs.] Both liquid Argon time projection chambers (LAr TPC) and water Cherenkov (WC) detectors have a much better spatial resolution, which at Icecube/DeepCore is eventually limited by the spacing between the Digital Optical Modules in the ice. 
\item[ Lower-energy sensitivity.] IceCube/DeepCore lacks sensitivity to showers with energies below a few GeV, while WC are a priori sensitive to the production of any charged particles as long as they lie above Cherenkov threshold. LAr TPCs are also sensitive to neutrino interactions with energies as low as 50~MeV~\cite{Acciarri:2014gev,Acciarri:2018myr}. Since the flux of atmospheric neutrinos rises steeply at low energies, the total number of events will be comparatively larger on a per-detector-mass basis. 
\item[ Availability of neutrino beam data.] Besides the atmospheric neutrino sample, at SK, HK and DUNE an additional event sample is available from the use of a conventional neutrino beam. In addition to atmospheric neutrino events, we also include the beam data in our study. 
\end{description}

The manuscript is structured as follows.
Our computation of the cross section for the interaction vertex is described in Sec.~\ref{sec:xsec} for both scenarios discussed above. Section~\ref{sec:decay} contains some relevant details regarding the lifetime of the HNL and the branching ratio into visible states, while the more technical details used in the computation of the signal event rates are discussed in Sec.~\ref{sec:simulation}. Section~\ref{sec:bg} contains an explanation of the expected background event rates, while we present our results in Sec.~\ref{sec:results} and summarize and conclude in Sec.~\ref{sec:conclusions}.

%%%%%%%%%%%%%%%%%%%%
\section {Cross section calculation}
\label{sec:xsec}
%%%%%%%%%%%%%%%%%%%%

In this section we describe our derivation of the interaction cross section in the quasi-elastic (QE) regime. Our calculation for the cross section in the deep inelastic scattering (DIS) regime is very similar to the one performed in Ref.~\cite{Coloma:2015kiu}, with the difference that in the present work the HNL mass has been properly included in the computation. 

We consider the scattering of a neutrino ($\nu_\ell$) or an anti-neutrino ($\bar{\nu}_\ell$) off a nuclear target and we denote by  $k=(E_\nu,\mathbf{k})$ and $k=(E^\prime_\nu,\mathbf{k}^\prime)$ the momentum of the initial and outgoing lepton.
The double-differential cross section for a neutral-current (NC) process can be written as~\cite{Shen:2012xz,Benhar:2006nr}
\begin{equation}
\Big(\frac{d\sigma}{dE^\prime_\nu d\Omega^\prime}\Big)_{\nu_\ell/\bar{\nu}_\ell}= \frac{G^2}{4\pi^2}\, k^\prime E^\prime\, L_{\mu \nu} R^{\mu \nu}\, ,
\label{eq:xsec_def}
\end{equation}
where $G=G_F \cos\theta_c$ with $\cos\theta_c=0.97425$~\cite{Nakamura:2010zzi} and 
for the Fermi coupling constant we use $G_F = 1.1803 \times 10^{-5}$ ~\cite{Herczeg:1999dx}.  
Performing the Lorentz contraction of the leptonic and hadronic tensor of Eq.~\eqref{eq:xsec_def}, we obtain
\begin{align}
\Big(\frac{d\sigma}{dE^\prime_\nu d\Omega^\prime }\Big)_{\nu/\bar{\nu}}&=\frac{G_F^2  \cos^2\theta_c}{4\pi^2}\frac{k^\prime}{2 E_\nu}\left[\hat{L}_{CC}R_{CC}+2\hat{L}_{CL}R_{CL}\right. \nonumber\\
&\left.+\hat{L}_{LL}R_{LL}+\hat{L}_{T}R_{T}\pm 2 \hat{L}_{T^\prime}R_{T^\prime}\right]\ .
\label{eq:cross_sec}
\end{align}
 We define $q$ and $Q$ as
\begin{align}
q&=k-k^\prime=(\omega, {\bf q})\, , \quad Q=k+k^\prime=(\Omega, {\bf Q}) \, .
\end{align}
Taking the three-momentum transfer along the $z$ axis and the total three-momentum in the $x-z$ plane, we write the leptonic kinematical factors as
\begin{align}
\hat{L}_{CC}&={\Omega}^2-q_z^2 -M_N^2,\ \ \ \ \ \ \ \ \hat{L}_{CL}=(-\Omega Q_z+\omega q_z)\nonumber\\
\hat{L}_{LL}&= {Q_z}^2 - \omega^2 + M_N^2,\ \ \ \ \ \ \ \hat{L}_{T}= \frac{{Q_x}^2}{2}- q^2 + M_N^2 \nonumber\\
\hat{L}_{T^\prime}&= \Omega q_z- \omega Q_z\, ,
\end{align}
with $M_N^2=k^{\prime\, 2}$ the mass of the outgoing neutrino.
The five electroweak response functions correspond to different components of the hadron tensor
\begin{align}
R_{CC}&=R^{00},\ \ \ \ \ \ \ \ R_{CL}=-\frac{1}{2}(R^{03}+R^{30})\nonumber\\
R_{LL}&=R^{33},\ \ \ \ \ \ \ R_T=R^{11}+R^{22}\nonumber\\
R_{T^\prime}&=-\frac{i}{2}(R^{12}-R^{21})\,.
\label{eq:response_w}
\end{align}
The hadronic tensor describes the transition between the initial and final nuclear states $|\Psi_0\rangle$ and $|\Psi_f\rangle$, with energies $E_0$ and $E_f$. For nuclei of spin-zero, it can be written as
\begin{align}
R^{\mu\nu}({\bf q},\omega)= \sum_f \langle \Psi_0|j^{\mu \, \dagger}({\bf q},\omega)|\Psi_f\rangle \langle \Psi_f| j^\nu({\bf q},\omega) |\Psi_0 \rangle \delta (E_0+\omega -E_f)\, .
\label{eq:had_tens}
\end{align}
where we sum over all hadronic final states and $J^\mu({\bf q},\omega)$ is the nuclear current operator. 
The one-body current operator is the sum of vector (V) and axial (A) terms NC processes and it can be written as
\begin{align}
&j^\mu= (J^\mu_V+J^\mu_A) \nonumber\\
&J^\mu_V={\cal F}_1 \gamma^\mu+ i \sigma^{\mu\nu}q_\nu \frac{{\cal F}_2}{2M}\nonumber\\
&J^\mu_A=-\gamma^\mu \gamma_5 {\cal F}_A \ .
\end{align}

The single-nucleon form factors relevant to the NC neutrino-nucleon scattering ${\cal F}^N_i$ (i=1,2) and $N=p,n$ read
\begin{align}
{\cal F}^N_i=&\pm \frac{1}{2}(F_i^p-F_i^n)-2 \sin^2\theta_W F^N_i\, ,\nonumber\\
{\cal F}_A=&\frac{1}{2}(F_A^s\pm F_A)\, ,
\end{align}
where the + (-) sign is for the proton (neutron), and $\theta_W$ is the Weinberg angle ($\sin^2\theta_W = 0.2312$~\cite{Nakamura:2010zzi}). 
We assumed a dipole parametrization for the axial form factor ${\cal F}_A=F_A$ 
\begin{align}
 F_A &=\frac{g_A}{( 1- q^2/ M_A^2 )^2}\ ,
\end{align}
where the nucleon axial-vector coupling constant is taken to be $g_A=1.2694$~\cite{PDG} and the axial mass 
$M_A=1.049$ GeV. 
The form factors
$F_A^s$ is introduced to describe the strangeness content of the nucleon. Following Ref.~\cite{Leitner:2008ue}, 
we set
\begin{align}
F_A^s&=-\frac{0.15}{(1-q^2/M_A^2 )^2}\, .
\end{align}

The second reaction mechanism we consider is the heavy neutrino production via a transition magnetic moment
\begin{align}
\Big(\frac{d\sigma}{dE^\prime_\nu d\Omega^\prime}\Big)_{\nu_\ell/\bar{\nu}_\ell}= \frac{1}{4\pi}\frac{\alpha\mu_{\rm tr}^2}{Q^4}\frac{E_\nu^\prime}{E_\nu}\, L_{\mu \nu} R^{\mu \nu}\, ,
\label{eq:diffTran}
\end{align}
where $\alpha=1/137$ is the electromagnetic fine structure constant, and $\mu_{\rm tr}=0.296 $ MeV$^{-1}$ is the neutrino transition magnetic moment. 
In this case we can write the contraction between the lepton and the hadron tensor as
\begin{align}
    L_{\mu \nu} R^{\mu \nu}=&\frac{W_1}{2}(2M_N^2-Q^2)(M_N^2+Q^2)+
    \frac{W_2}{2}\big[ E_\nu^2(Q^2-3M_N^2)+2 E_\nu E^\prime_\nu(M_N^2+Q^2) \nonumber\\
    & +(E_\nu^\prime-M_N)(E_\nu^\prime+M_N)(M_N^2+Q^2)\big]
\end{align}
where
\begin{align}
    W_1&=\frac{(R^{11}+R^{22})}{2} \nonumber\\
    W_2&=\frac{Q^4}{|{\bf q}|^4}R^{00}+\frac{Q^2}{|{\bf q}|^2}\frac{(R^{11}+R^{22})}{2}
\end{align}
with $M_N$ the mass of the right handed neutrino. 
The general expression of the hadron tensor is given in Eq.~\eqref{eq:had_tens}. Since the exchanged boson is a $\gamma$, for the hadron vertex, the current operator coincides with the electromagnetic one
\begin{align}
j^\mu_{\rm EM}={\mathcal F}_1 \gamma^\mu+ i \sigma^{\mu\nu}q_\nu \frac{{\mathcal F}_2}{2m_N}\, .
\label{rel:1b:curr}
\end{align}
The isoscalar (S) and isovector (V) form factors, ${\mathcal F}_1$ and ${\mathcal F}_2$,
are given by combination of the Dirac and Pauli ones, $F_{1}$ and $F_2$, as
\begin{align}
{\mathcal F}_{1,2}= \frac{1}{2}[F_{1,2}^S+F_{1,2}^V\tau_z]
\end{align}
where $\tau_z$ is the isospin operator and 
\begin{align}
F_{1,2}^S=F_{1,2}^p + F_{1,2}^n,\ \ \ \ \ \ \ \ F_{1,2}^V=F_{1,2}^p - F_{1,2}^n\, . 
\label{f12:iso}
\end{align}

\subsection*{Spectral function formalism}
The cross sections computed in this work have been obtained using a factorization scheme. For large values of the energy transfer $ \omega$ and momentum transfer $|{\bf q}|$, we can factorize the hadronic final state and provide a relativistic description of $|\Psi_f\rangle$ and of the current operator. 
To model the dynamics of the target nucleus and account for correlation effects we used a realistic spectral function obtained from the non-local dispersive optical model (DOM)~\cite{Mahaux91,Mahzoon:2014}. 

%%%%%%%%%%%%%%%%%%%%%%
\begin{figure*}[t!]
\begin{center}
\includegraphics[width=.41\columnwidth]{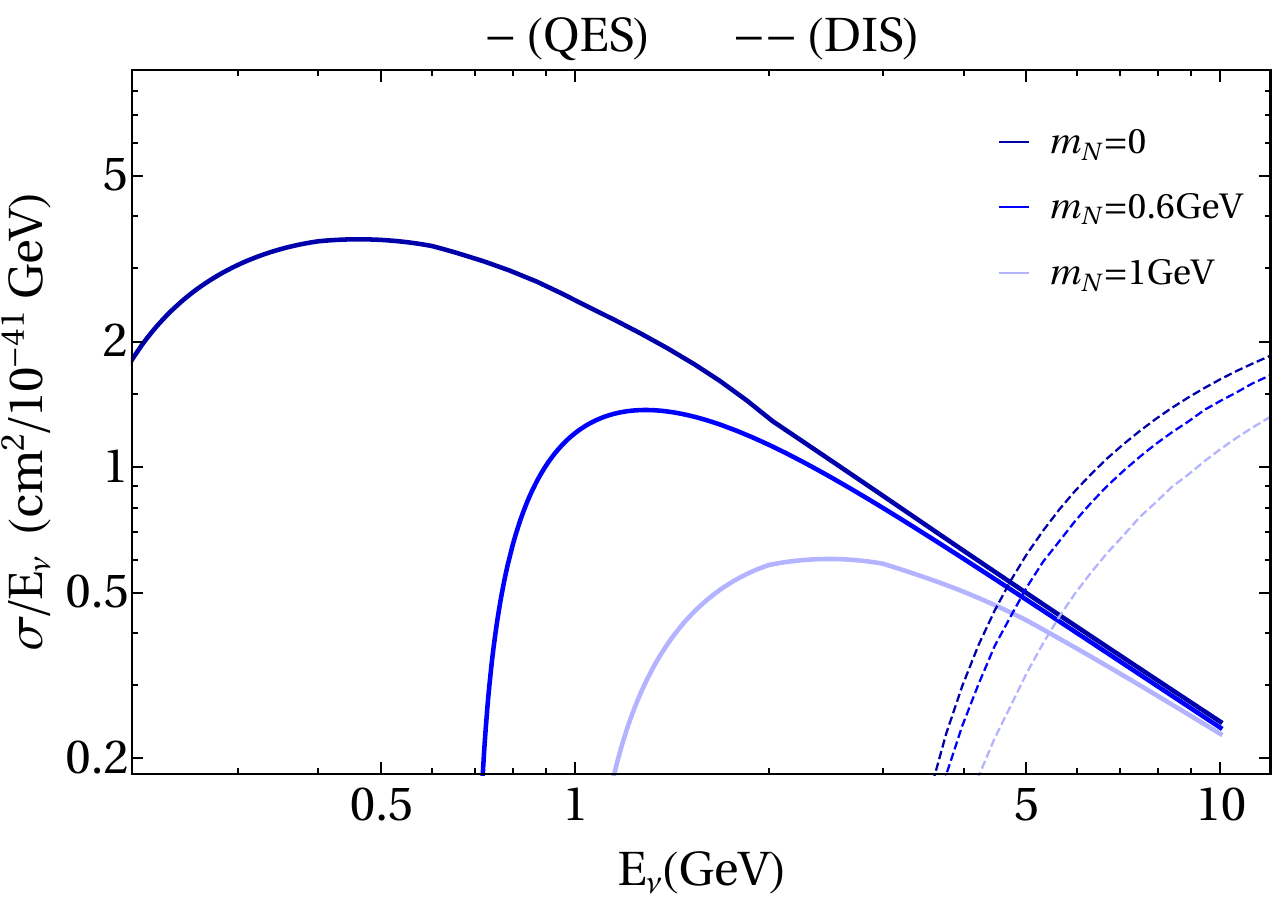}
\includegraphics[width=.42\columnwidth]{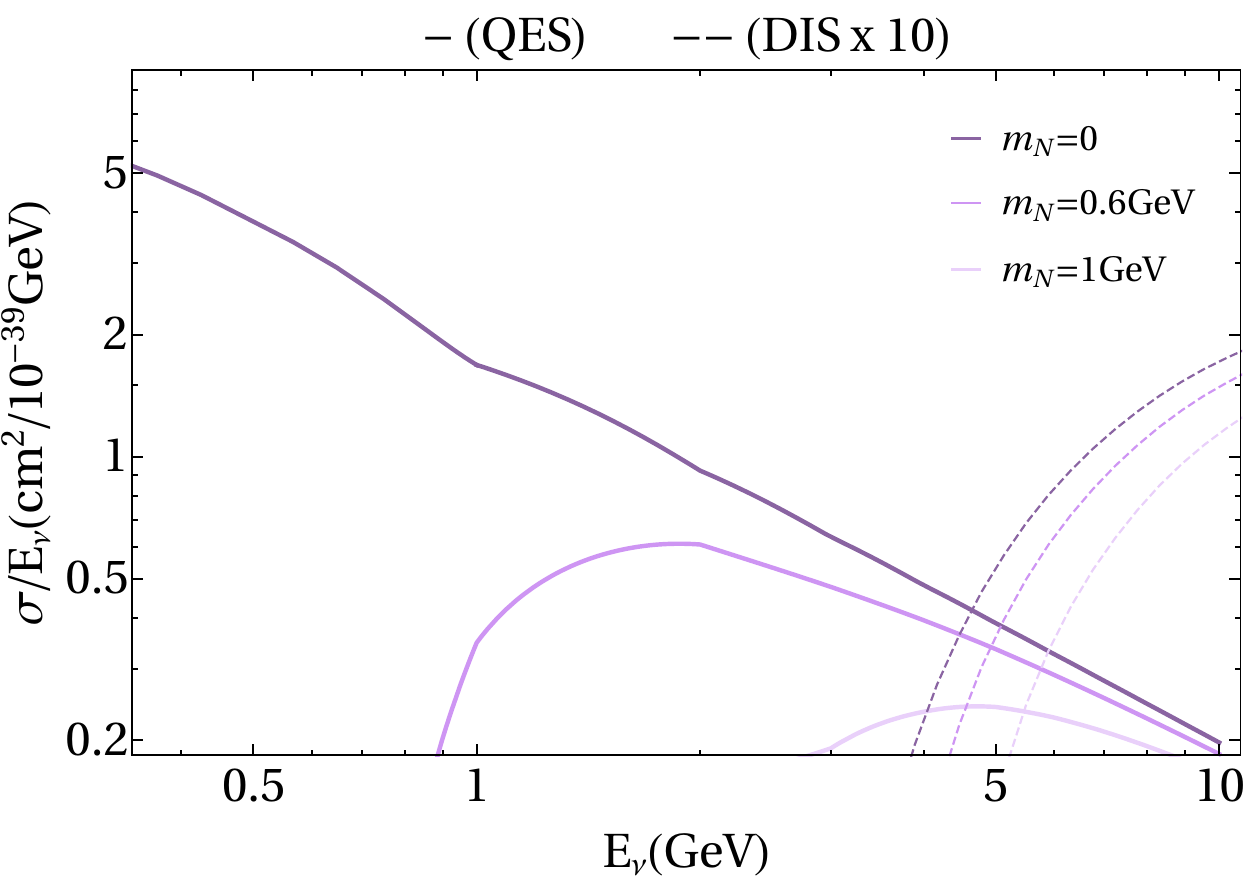}
\caption{Here we display the cross sections in the mass-mixing (left) and transition magnetic moment (right) cases. The QE (solid) and DIS (dashed) contributions are shown separately for representative values of the HNL neutrino mass. Note that for the magnetic moment case, the DIS contribution has been magnified by 10 times in order for it to be visible. }
\label{fig:cx}
\end{center}
\end{figure*}
%%%%%%%%%%%%%%%%%%%%%%

We consider only the one-body current terms and we rewrite the hadronic final state as 
\begin{equation}
|\Psi_f\rangle \rightarrow |p\rangle \otimes |\psi_f^{A-1}\rangle
\label{fact:ans}
\end{equation}
where the outgoing nucleon with momentum $|{\bf p}|$ and energy $e({\bf p})=\sqrt{|{\bf p}|^2+m_N^2}$ is described in terms of a plane wave, while $|\psi_f^{A-1}\rangle$ stems for the \hbox{$(A-1)$-body} spectator system. 
The energy and momentum of the latter are obtained by energy and momentum conservation relations
\begin{align}
E_f^{A-1}=\omega +E_0-e({\bf p})\, ,\quad {\bf P}^{A-1}_f={\bf q}-{\bf p}\, .
\end{align}
The hadron tensor of Eq.~\eqref{eq:had_tens} can be easily rewritten by inserting a single-nucleon completeness relation
\begin{align}
R^{\mu\nu}=& \int \frac{d^3k}{(2\pi)^3} dE S_h({\bf k},E)\frac{m^2}{e({\bf k})e({\bf k+q})} \sum_{i}\, \langle k | {j_{i}^\mu}^\dagger |k+q \rangle \langle k+q |  j_{i}^\nu | k \rangle \delta(\tilde{\omega}+e({\bf k})-e(\mathbf{k+q}))\,,
\label{had:tens}
\end{align}
where $m$ is the nucleon nucleon mass. 
In order to account for the off-shellness of the initial nucleon, we introduced $\tilde{\omega}=\omega-E+m-e({\bf k})$ and replace $q=(\omega, {\bf q})\rightarrow  \tilde{q}=(\tilde{\omega}, {\bf q})$ in the calculation of the hadronic tensor.
The factors $m/e({\bf k})$ and $m/e({\bf k+q})$ ensure the implicit covariant normalization of the nucleon quadri-spinors.

The spectral function, $S_h({\bf k},E)$ in Eq.~\eqref{had:tens}, is obtained for $^{16}$O using the DOM. This method constrains a complex self-energy $\Sigma^*$ using both scattering and bound-state data of $^{16}$O~\cite{Mahaux91,Mahzoon:2014}. The self-energy is a complex, nonlocal, energy-dependent potential that unites the nuclear structure and reaction domains through dispersion relations~\cite{Mahaux91,Mahzoon:2014,Dickhoff:2017}. The Dyson equation generates the single-particle propagator, or Green's function, $G_h({\bf k},E)$ from which bound-state and scattering observables can be deduced~\cite{Atkinson:2020}.
The energy dependence of the self-energy ensures that many-body correlations manifest in $G_h({\bf k},E)$, providing a description beyond that of a mean field.
The spectral function determined by $G_h({\bf k},E)$ in the following equation 
\begin{equation}
   S_h({\bf k},E)=\frac{1}{\pi}\textrm{Im}G_h({\bf k},E)
\label{eq:spectral_function}
\end{equation}
yields the probability density of finding a nucleon in the target nucleus with a given momentum ${\bf k}$ and removal energy $E$.

\begin{figure*}[]
\begin{center}
\includegraphics[scale=0.85]{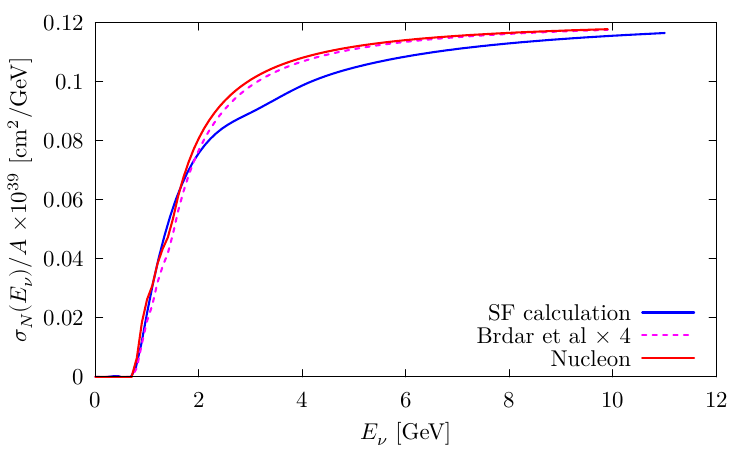}
\caption{Total cross section for heavy neutrino production via magnetic moment transition, as a function of the incoming neutrino energy. The solid red curve corresponds to the scattering on a single nucleon at rest. The blue line displays the nuclear cross section on a $^{16}$O target computed within the SF formalism, divided it by the number of nucleons. As a cross check, we have also computed the cross section as in Ref.\cite{Brdar:2020quo} (dashed magenta line), replacing the electron mass with the nucleon one and integrating over the recoil energy (note that the results obtained following Ref.~\cite{Brdar:2020quo} have been multiplied by a factor 4 to account for the different convention used in the definition of the magnetic moment). }
\label{fig:cx:nucleon}
\end{center}
\end{figure*}

In Fig.~\ref{fig:cx} we show representative total cross sections divided by the incoming neutrino energy in both the mass-mixing with $|U_{\tau 4}|^{2} =10^{-2}$ (left panel) and magnetic moment with $\mu = 10^{-8}~\mu_{B}$ (right panel), where $\mu_{B} \equiv e/(2 m_{e})$ stands for the Bohr magneton. The cross sections displayed in both panels have been computed for an $^{16}$O target, using the corresponding nuclear SF. 
The difference between the neutrino scattering cross section on a free nucleon at rest and the one on a nuclear target (where protons and neutrons are bound and subject to Fermi motion) divided by the number of nucleons $A$, can be best appreciated in Fig.~\ref{fig:cx:nucleon} by comparing the solid red and blue curves, respectively. The total cross section per nucleon for the magnetic dipole coupling and a massive neutrino with $m_N= 0.6$ GeV is plotted as a function of the incoming neutrino energy $E_\nu$. The results obtained using the $^{16}$O SF are quenched with respect to the free nucleon case, and nuclear effects become less visible for large values of $E_\nu$. As a cross check, we have also computed the cross section as in Ref.\cite{Brdar:2020quo}, replacing the electron mass with the nucleon one and integrating over the recoil energy (dashed magenta curve). The results obtained following Ref.\cite{Brdar:2020quo} have been multiplied by a factor 4 to account for the different convention used in the definition of the magnetic moment.
We find a very good agreement with our solid line in Fig. ~\ref{fig:cx:nucleon}, with only minor differences at low energies due to the terms proportional to $\mathcal{F}_2$, which have been neglected in Ref.~\cite{Brdar:2020quo} and included in our results.  

%%%%%%%%%%%%%%%%%%%%%%
%%%%%%%%%%%%%%%%%%%%%%

\subsection{DIS regime}

HNL can be produced via mixing with the active neutrinos. In that scenario, which described by Eq.~\ref{eq:seesaw}, the interaction of the active neutrinos inside the detector at high energies ($E\leq 10$~GeV) is mainly described in the DIS regime. At very high energies, neutrinos are able to resolve inside the nucleons and the interaction happens with the quarks. Similar to the quasi-elastic regime, the double-differential cross section can be written as the product of a leptonic and a hadronic tensor. In the limit where the momentum transfer ($q$) is much smaller than the mass of the Z boson ($q << M_{z}$), the amplitude square of this process is given by 
\begin{equation}
  |M|^2 = \frac{e^4}{16 s^4_{w} c^4_{w}M^4_{z}} L_{\alpha\beta} W^{\alpha\beta}
\end{equation}
where $e$ is the electric charge. $s_{w} = \sin\theta_{w}$ and $c_{w} = \cos\theta_{w}$ are the sine and cosine of the weak mixing angle. The tensor $L_{\alpha\beta}$ comes from the coupling to the lepton current. The leptonic tensor is given by QED and it takes the following form
\begin{equation}
  L^{\alpha\beta} = k^{\alpha}k'^{\beta} + k^{\beta}k'^{\alpha} - g^{\alpha\beta} k\cdot k' + \imath \epsilon^{\alpha\beta\rho\eta}k^{\rho}k'^{\eta}
\end{equation}
where, using the same nomenclature as in the quasi-elastic regime, $k$ and $k'$ denote the initial and outgoing leptonic momentum. The hadronic current is described by the tensor $W^{\alpha\beta}$ and it takes the general form
\begin{align}
  W_{\alpha\beta} &= - \frac{W_{1}}{M}\left(g_{\alpha\beta} - \frac{q_{\alpha}q_{\beta}}{q^2}\right) + \frac{W_{2}}{M^{2}(E_{\nu} - E'_{\nu})}\left(P_{\alpha} - \frac{q\cdot P q_{\alpha}}{q^2}\right)\left(P_{\beta} - \frac{q\cdot P q_{\beta}}{q^2}\right)\\\nonumber
  &+\imath\frac{W_{3}}{2M^2(E_{\nu} - E'_{\nu})}\epsilon_{\alpha\beta\rho\eta} P^{\rho}q^{\eta}
\end{align}
where, similar as before, $W_{i}$ correspond to the dimensionless structure functions. M is the nucleon mass. For an isoscalar nucleon ($N= 1/2(n+p)$), the structure relations takes the form~\cite{Fogli:1986iu,Quigg:2013ufa} 
\begin{align}
  W_{2} = 2x \left\{((c^u_{L})^2 + (c^u_{R})^2 + (c^d_{L})^2 + (c^d_{R})^2)[u(x) + \bar{u}(x) + d(x) + \bar{d}(x)]\right\}\\
  xW_{3} = 2x \left\{((c^u_{L})^2 - (c^u_{R})^2 + (c^d_{L})^2 - (c^d_{R})^2)[u(x) - \bar{u}(x) + d(x) - \bar{d}(x)]\right\}
    \label{eq:formfactors} 
\end{align}
In the structure functions, we include the contribution of the first generation of quarks, that we expect that dominates the cross section, at least at lower energies. $u(x)$, $d(x)$, $\bar{u}(x)$ and $\bar{d}(x)$ correspond to the quark distribution functions inside the nucleons. For $W_{1}$, we can use the Callan-Gross relation $2x W_{1} = W_{2}$, where $x = Q^2/2M(E_{\nu} - E'_{\nu})$. The coefficients $c^u_{L} = 1/2 - 2/3s^2_{w}$, $c^u_R = -2/3 s^2_{w}$, $c^d_{L} = -1/2 + 1/3 s^2_{w}$, $c^d_R = 1/3 s^2_{w}$ are the vertex couplings between the quarks and the Z boson. Taking the product between the leptonic and hadronic tensors, we get the following
\begin{align}
  &L_{\alpha\beta} W^{\alpha\beta} = -\frac{W_{1}}{M Q^2}(M^2_{N} - 2Q^2)(M^2_{N} + Q^2) - \frac{W_{3}}{M (E_{\nu} - E'_{\nu})}(M^2_{\nu} (E_{\nu} - E'_{\nu}) - Q^2(E_{\nu} + E'_{\nu}))\\\nonumber
  &+ \frac{W_{2}}{Q^4(E_{\nu} - E'_{\nu})}(Q^4(4E_{\nu}E'_{\nu} - M^2_{N}) - M^2_{N}Q^2(E_{\nu} - E'_{\nu})(3E_{\nu} + E'_{\nu}) + M^4_{N}(E_{\nu} - E'_{\nu})^2 - Q^6)
    \label{eq:tensorproductDIS}   
\end{align}
In the limit where $m_{N} =0$, we recover the standard expression for a NC interaction in the DIS regime.

The heavy sterile neutrinos can also be produced via a transition dipole moment, Eq.~\ref{eq:dipole}. In this scenario, the neutrino-nucleon interaction is mediated by a photon. At very high energies, the photon interact with the quarks inside the nucleons. The double differential cross section is given by Eq.~\ref{eq:diffTran}. Beyond the fact that the mediator in this scenario is massless, the main difference with the mixing scenario lies in the leptonic tensor, that in dipole moment scenario is given by the effective operator Eq.~\ref{eq:dipole} and takes the form
\begin{align}
  L_{\mu\nu} &= tr[(\cancel{k'} - m_{N})[\cancel{q},\gamma^{\mu}]P_{L}\cancel{k}[\gamma^{\nu},\cancel{q}]P_{R}]
\end{align}
Since the dipole interaction is parity conserving, the hadronic tensor only contains symmetric terms. The product between the leptonic and hadronic tensors gives
\begin{align}
  L_{\alpha\beta} W^{\alpha\beta} &= \frac{W_{1}}{2}(2m^2_{N} - Q^2)(m^2_{N} + Q^2)\\\nonumber
  &+ \frac{W_{2}}{2}(E^2_{\nu}(Q^2 - 3m^2_{N}) + 2E_{\nu}E'_{\nu}(m^2_{N} + Q^2) + (E'_{\nu} - m_{N}) (E'_{\nu} + m_{N}) (m^2_{N} + Q^2))
\end{align}

%%%%%%%%%%%%%%%%%%%%%%%%%%
\section{HNL decay}
\label{sec:decay}
%%%%%%%%%%%%%%%%%%%%%%%%%%

\subsection{Mixing with tau neutrinos}

The computation of the lifetime of the HNL in this scenario, as well as the branching ratios into the different decay channels, is done following Ref.~\cite{Coloma:2020lgy}. For a HNL with mass $m_N \sim \mathcal{O}(\rm{GeV})$ that mixes primarily with $\nu_\tau$, the decay channels involving tau leptons are kinematically forbidden. In this mass range the HNL will primarily decay into two-body final states, namely, neutral mesons plus a tau neutrino. While $N \to \nu_\tau \pi^0$ is the dominant decay channel of this type (with a branching ratio between 20\% and 30\%, depending on $M_N$), non-negligible branching ratios (sometimes as large as 10\%) are also found for $N\to \nu_\tau \rho$ and $N\to \nu_\tau \eta$ (see bottom panels in Fig.~1 in Ref.~\cite{Coloma:2020lgy}). Slightly smaller branching ratios are found for the decay into the three-body fully leptonic decay channels, $N\to \nu_\tau e e$ and $N \to \nu_\tau \mu\mu$. The resulting decay length in the lab frame, for a neutrino with energy $E_N$ and total decay width $\Gamma_{tot}$, is in the ballpark
\be 
L_{{\rm lab}} = c \tau \gamma \beta  \simeq 10~{\rm m}~\left(\frac{10^{-2}}{|U_{\tau 4}|^{2}}\right)~\left(\frac{0.5~{\rm GeV}}{m_N}\right)^{5}~\left(\frac{E_{N}}{1~{\rm GeV}}\right) \, ,
\label{eq:LengthHNL}
\ee
where $\tau = 1/\Gamma_{tot}$ is the lifetime of the HNL in its rest frame. 

\begin{figure}[t!]
\begin{center}
\includegraphics[scale=0.85]{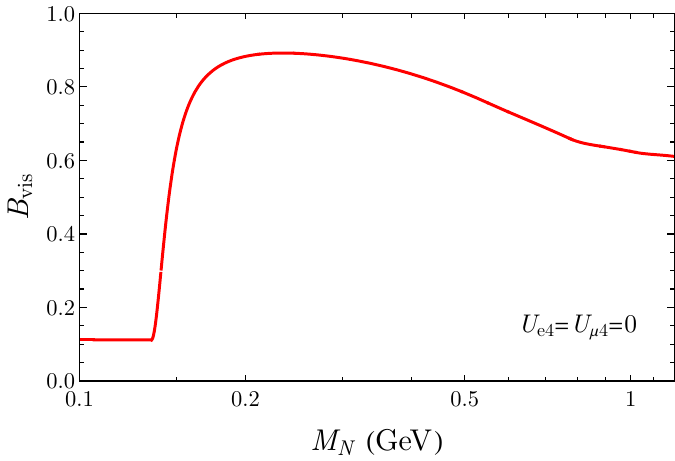}
\caption{Branching ratio for the decay of a HNL of mass $M_N$ into visible final states, defined in Eq.~\ref{eq:BRvis-mixing}, assuming it only mixes with tau neutrinos. }
\label{fig:BRvis}
\end{center}
\end{figure}

When computing the event rates, however, we take into account that in this range of masses the branching ratio for the decay $N\to \nu_\tau\nu_\alpha\nu_\alpha$ is dominant and would lead to an invisible HNL decay in any neutrino detector. In order to take this into account, our event rates are multiplied by the branching ratio into visible decay channels 
\be
\label{eq:BRvis-mixing}
\mathcal{B}_{vis} = 1 - \frac{\Gamma (N\to\nu\nu\nu)}{\Gamma_{tot}} \, .
\ee
Figure~\ref{fig:BRvis} shows the branching ratio into visible states for a HNL that mixes with tau neutrinos only, as a function of its mass.

%%%%%%%%%%%%%%%%%%%%%%
%%%%%%%%%%%%%%%%%%%%%%

\subsection{Magnetic Dipole Moment}

In the case of the magnetic dipole coupling, for simplicity we will assume that the mixing to light sterile neutrinos is negligible. In this case, the operator in Eq.~\eqref{eq:dipole} induces the two-body decay $N \to \nu \gamma$ with the width 
\be
\Gamma(N \rightarrow \nu \gamma) = \frac{\mu^{2}m_N^{3}}{4 \pi} \, ,
\ee
which is the only relevant decay channel in this scenario. Thus, in this case we have $\mathcal{B}_{vis} = 1$, and the decay length in the lab frame reads
\be 
L_{{\rm lab}} = c \tau \gamma \beta  \simeq 2.5~{\rm m}~\left(\frac{10^{-8}~\mu_{B}}{\mu}\right)^{2}~\left(\frac{100~{\rm MeV}}{m_N}\right)^{4}~\left(\frac{E_{N}}{1~{\rm GeV}}\right) \, .
\label{eq:lengthdipole}
\ee

%%%%%%%%%%%%%%%%%%
\section{Calculation of Double-Bang Event Rates}
\label{sec:simulation}
%%%%%%%%%%%%%%%%%%

The Double-Bang (DB) signal consists on a composite signal given by the interaction of an active neutrino (``first bang''), where a HNL is produced, and the posterior decay of the heavy state (``second bang'') after the propagation of a macroscopic distance. The separation between both bangs depends on the mass of the heavy state and the decay channel into SM particles. As stated before, our goal is to determine the present and future sensitivities to this signal in two of the most relevant type of detectors for neutrino physics: water Cherenkov and Liquid Argon TPC. In particular, we will consider the present sensitivity of Super-Kamiokande (SK) and its future upgrade Hyper-kamiokande (HK), as well as DUNE. As a neutrino source, we consider the two largest fluxes that can be measured by those experiments: the atmospheric neutrino flux and the neutrinos generated in an accelerator. In the case of SK and HK, the beam experiments are called T2K and T2HK respectively. 

Note that the rate of NC $\nu_{\tau}$ events at each experimental setup sets the scale for their ability to probe HNLs. For illustration, we plot in Fig.~\ref{fig:events} number of $\nu_{\tau}$ induced NC events at Super-K via the atmospheric flux, and T2K and DUNE via their respective beam fluxes. From this we expect the existing sensitivity of Super-K and the future DUNE beam sensitivity to set the most stringent limits on HNLs.

For beam-induced neutrino events, the DB rate generated by the interaction of $\nu_{\alpha}$ which upscatter into the heavy state, for a HNL with lifetime $c\tau$, is proportional to 
\be
N^\alpha_{\rm DB}(c\tau) = \sum_{\beta}  \mathcal{B}_{vis} \!\int \! dE_\nu dc_{\theta}dE_N \frac{d \phi_{\nu_\beta}}{dE_\nu}  P_{\beta\alpha}(E_\nu)~\frac{d^2 \sigma_{\nu_{\alpha}N}}{dE_N dc_{\theta}}(E_{\nu})~P_{d}(c\tau,E_N)~V_{\rm det} (c\tau,E_N, c_{\theta}),
\label{eq:events}
\ee
where $E_\nu$ is the incident neutrino energy, while $E_N$ and $c_{\theta}\equiv \cos\theta$ refer to the energy and the scattering angle of the heavy state with respect to the incident neutrino direction. The flux of active neutrinos is given by $\phi_{\beta}$ and it will have oscillated into $\nu_{\alpha}$ by the time that they arrive to the detector, with a probability $P_{\beta\alpha}$. The cross section for neutrino upscattering into the heavy state is given by $\sigma_{\nu_{\alpha}N}$. The NC quasi-elastic neutrino interaction is computed as described in Sec.~\ref{sec:xsec}, for the DIS regime, we followed Ref.~\cite{Coloma:2017ppo} including the correct dependence with the HNL mass. After propagation over a distance $L$, the HNL will decay with a probability $P_{d}=e^{-L/L_{\rm lab}}/L_{\rm lab} \equiv P_{d}(c\tau, E_N)$, where $L_{\rm lab}$ depends on the specific model considered as outlined in Sec.~\ref{sec:decay}. Finally, the effective volume $V_{\rm det}$ includes the probability that the second bang takes place inside the detector, which depends on the expected separation between the two bangs for a given value of $c\tau$ and the boost factor. In the case of atmospheric neutrinos, the number of events in Eq.~\eqref{eq:events} should be modified to to take into account the dependence of $P_{\beta\alpha}$, $\phi_{\nu_\beta}$, and $V_{\rm det}$ on the solid angle corresponding to the neutrino trajectory before the scattering, which would also have to be integrated over.

In the case of T2K, we use the fluxes published by the collaboration~\cite{Abe:2021gky}. A similar energy-dependent flux will be used for T2HK, although we will assume a beam power of 1.3~MW~\cite{Hasegawa:2018bjo,Igarashi:2019txx} (see also e.g., Ref.\cite{Ballett:2016daj}), corresponding to $2.7\times 10^{22}$~POT in 10 years. For DUNE, we will use the collaboration flux predictions~\cite{Abi:2020evt}, and a beam power of 1.2~MW that will correspond to $1.1\times 10^{22}$~POT in 10 years. While for beam calculations we consider $\nu_\mu$ as the initial neutrino flavor, neglecting the intrinsic contamination of the beam, for atmospheric neutrinos we consider both $\beta = e, \mu$, according to Honda's table~\cite{Honda:2015fha}. In the case of atmospheric neutrinos, the flux and the oscillation probability also depend on the zenith angle. We will integrate over all the incoming neutrino directions and separations between the ``bangs'' to obtain the total DB rate. Also, note that antineutrino events will give a similar contribution to the total number of events, replacing $ \phi_{\nu_\mu}$, $\sigma_{\nu_\tau N}$ and $P_{\mu\tau}$  in Eq.~\eqref{eq:events} by their analogous expressions for antineutrinos.

To be detected, a DB signal must happen inside the detector. The asymmetry of the detectors in each direction implies that the effective volume of the detector will depend on the separation between the bangs, but also on the scattering angle. The effective detector volume has been obtained via Monte Carlo integration, where we have included the detector geometry and the minimum separation between bangs that can be resolved in each detector. In the case of SK, the vertex resolution is 50~cm at 12.5~MeV and closer to 80~cm at 5~MeV~\cite{Abe:2016nxk}. In our simulation, we have conservatively assumed a minimum separation of 80~cm between the two events, for all neutrino energies. In the case of Liquid Argon TPC, the vertex resolution can be reduced to the millimeter scale~\cite{Arguelles:2019xgp}. However, in this case backgrounds can arise from several sources, as described in more detail in Sec.~\ref{sec:bg}. Thus, we have conservatively set the minimum separation between the two bangs at 20~cm.

%%%%%%%%%%%%%%%%%%%%%%%%%%%
\section{Expected backgrounds}
\label{sec:bg}
%%%%%%%%%%%%%%%%%%%%%%%%%%%

As argued in Ref.~\cite{Coloma:2017ppo}, the main advantage of searching for new physics through double-bang events is that the number of background events from SM processes is expected to be very small. Here we provide some estimates regarding the expected background rate for LAr TPC and WC detectors, since there are some relevant differences with respect to the case of Icecube/DeepCore. 

If the separation between the two events is above a few meters, the only background to this search comes from coincidental events, that is, two events taking place within the same time window $\Delta t$. Naively, for an atmospheric neutrino experiment the event rates can be estimated as in Ref.~\cite{Coloma:2017ppo}:
\be\label{eq:bg-atm}
R_{atm} \sim \frac{N_{atm} (N_{atm} - 1)}{2} \left(\frac{\Delta t }{T}\right)^2  ,
\ee 
where $N_{atm}$ is the number of atmospheric neutrino events which take place within a time period $T$. Our computation of the number of atmospheric events in each detector gives $N_{atm}^{DUNE} \simeq 1.2 \times 10^{3}~{\rm yr}^{-1}$,  $N_{atm}^{SK} \simeq 8 \times 10^{2}~{\rm yr}^{-1}$, and $N_{atm}^{HK} \simeq 6.6 \times 10^{3}~{\rm yr}^{-1}$. Thus, assuming $\Delta t = 10$ s (which is likely very conservative)  
 we find
$R_{atm}^{\rm DUNE} = 7.2\times 10^{-8}~{\rm yr}^{-1}$ ,
$R_{atm}^{\rm SK} = 3.2\times 10^{-8}~{\rm yr}^{-1}$,   and
$R_{atm}^{\rm HK} = 2.2\times 10^{-6}~{\rm yr}^{-1} $ for DUNE, SK and HK, respectively. 

In the case of the beam, the number of coincidental events within the same beam spill can be estimated in a similar fashion. In this case, the much smaller time window, which is given by the beam spill width (for example, $\Delta t = 10~\mu \textrm{s}$ in the case of DUNE~\cite{Abi:2018dnh}), leads to a negligible background level, well below that of atmospheric neutrinos. We find this to be the case for the coincidence of two beam events within the same spill, as well as for the case where one event comes from the beam and the other is an atmospheric neutrino.

However, once the separation between the two events is smaller than $\sim$50~cm additional backgrounds arise. For example, events in which a neutron is knocked out of the nucleus can exhibit a double-bang-like event topology, depending on the energy-dependent mean free path of the neutron. Similarly, photons can take some time to convert to electron/positron pairs and therefore NC events with vertex activity and a $\pi^0$ in the final state could {\it a priori} constitute a background as well (for example, the typical conversion distance for the two photons coming from a $\pi^0$ decay in LAr is 24~cm~\cite{Adams:2018sgn}). Thus, in our analysis we impose a minimum separation between the two events of 20~cm for the DUNE detector. 

At high energies ($\geq 5$~GeV), the hadronic cascade can also leave a double-bang event topology due to, e.g., the punch-through of neutral hadrons created at the neutrino interaction~\cite{CCFR:1992ajg}. This type of background becomes negligible for cascades separation on the order of $\sim 10$~m: thus, from Eqs.~\eqref{eq:LengthHNL} and~\eqref{eq:lengthdipole} we can see that this will hardly have an impact on the HNL scenario, while it may in principle affect our sensitivities to a magnetic dipole moment. However, the contribution from this background to the total event rates will be suppressed by the high energy tail of the neutrino flux and therefore we do not expect this source of background to have a large impact on our results. A proper study of the backgrounds to this search can only be addressed by the experimental collaborations, and lies outside of the scope of this work.

%%%%%%%%%%%%%%%%%%%%%%%%
\section{Results}
\label{sec:results}
%%%%%%%%%%%%%%%%%%%%%%%%

As a rough estimate of the DB sensitivity, one can expect the number of $\nu_\tau$-induced DB events to scale as 
\be 
N^\tau_{DB} \simeq N_{\nu_{\tau}}^{NC}(E_{\nu} \gtrsim m_{N})  \times \frac{V_{{\rm det}}}{V_{0}}
\frac{\sigma_{\nu_\tau N}}{\sigma^{NC}_{SM}} 
\times \mathcal{B}_{vis} \times P_{d} \, ,
\label{eq:est}
\ee
where $N_{\nu_{\tau}}^{NC}(E_{\nu} \gtrsim m_{N})$ is the number of NC $\nu_{\tau}$ events with energies above the kinematic threshold for $\nu_\tau \rightarrow N$ up-scattering, $\sigma^{NC}_{SM}$ is the neutrino NC interaction cross section in the SM, and $V_{0}$ is the full (fiducial) volume of the detector.

\begin{figure}[t!]
\begin{center}
\includegraphics[scale=0.50]{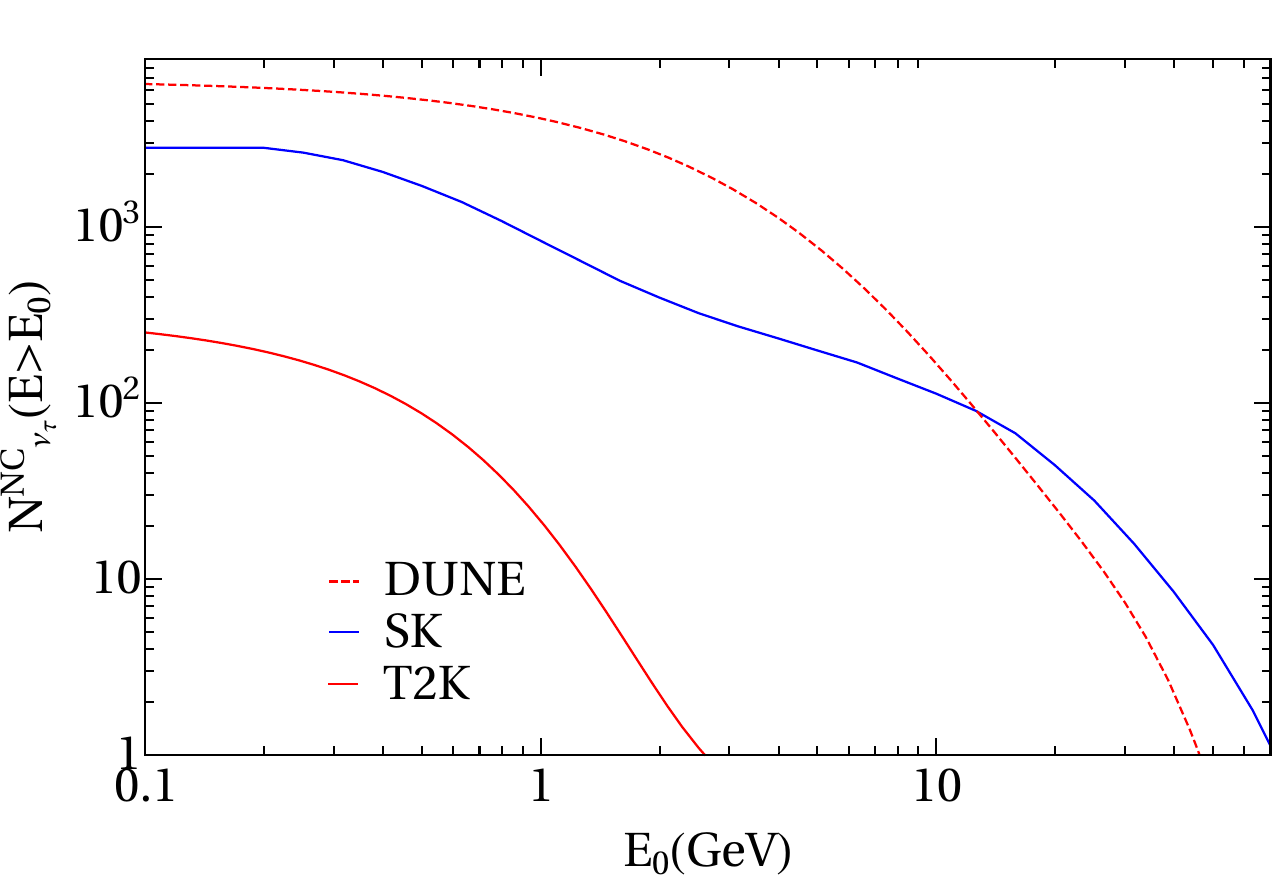}
\caption{Total number of standard NC $\nu_\tau$ interactions with neutrino energies above a given value $E_0$, for the atmospheric flux at SK, as well as for the T2K and for the DUNE beams. }
\label{fig:events}
\end{center}
\end{figure}

\begin{figure}[t!]
\begin{center}
\includegraphics[width=.95\columnwidth]{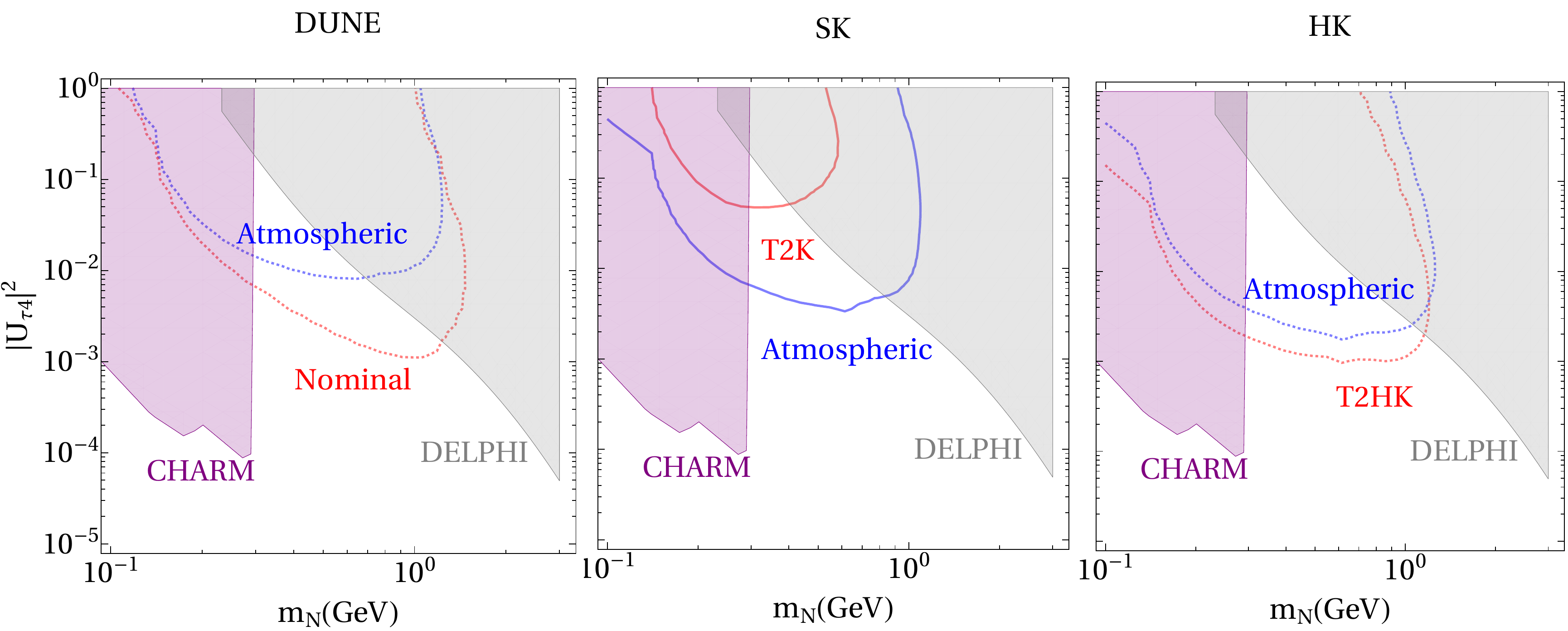}
\caption{Iso-contours indicating the regions where more than one DB event is expected, for the HNL scenario via mixing with $\nu_\tau$. Results are shown for DUNE (left), Super-K (center), and Hyper-K (right), for the atmospheric (blue) and nominal beam (red) fluxes separately. The solid curves for the Super-K and T2K sensitivity correspond to current exposures. The shaded regions indicate the regions of the parameter space disfavored by CHARM~\cite{Orloff:2002de} and DELPHI~\cite{Abreu:1996pa} data, taken from Ref.~\cite{Atre:2009rg}. 
Here we show direct experimental bounds on this scenario, obtained from searches for the HNL decay products; however, note that the mixing also faces indirect constraints (such as those derived from lepton flavor universality measurements, see e.g.~Ref.~\cite{Cvetic:2017gkt}). 
}
\label{fig:HNL}
\end{center}
\end{figure}

Let us first examine the mass-mixed HNL case at DUNE via the beam flux. For illustration, let us assume $m_{N}$ = 0.5 GeV. From our calculations, we find that the number of $\nu_{\tau}$-induced NC events above the kinematic threshold is $N_{\nu_{\tau}}^{NC}(E_{\nu} \gtrsim m_{N}) \simeq 4\times 10^{3}$, see Fig.~\ref{fig:events}. 
%%%%%%%%%%%%%%%%%%%%%%
%%%%%%%%%%%%%%%%%%%%%%
Next, inspection of Fig.~\ref{fig:cx} reveals that the total up-scattering cross section at this HNL mass is reduced by about a factor of $\sim (1/3) |U_{\tau 4}|^{2}$ compared to the SM case. Thus Eq.~(\ref{eq:est}) implies that we should expect sensitivities to the mixing angle, $|U_{\tau 4}|^{2} \gtrsim 10^{-3}$ at best (of course, the visible branching ratio, effective detector volume, and probability to decay will reduce this sensitivity somewhat). Our numerical results for the mixing scenario are shown in Fig.~\ref{fig:HNL} for DUNE, Super-K, and Hyper-K sensitivity in Fig. ~\ref{fig:HNL}. Upon inspection of the DUNE sensitivity in the left panel of Fig.~\ref{fig:HNL}, we see that the more detailed estimate finds that $|U_{\tau 4}|^{2} \gtrsim 3\times 10^{-3}$ can be reached at $m_{N} = 0.5$ GeV. One can also see that this is consistent with the requirement that the decay occurs within the detector volume, see Eq.~\eqref{eq:LengthHNL}. 
%%%%%%%%%%%%%%%%%%%%%%
%%%%%%%%%%%%%%%%%%%%%%

\begin{figure}[t!]
\begin{center}
\includegraphics[width=.95\columnwidth]{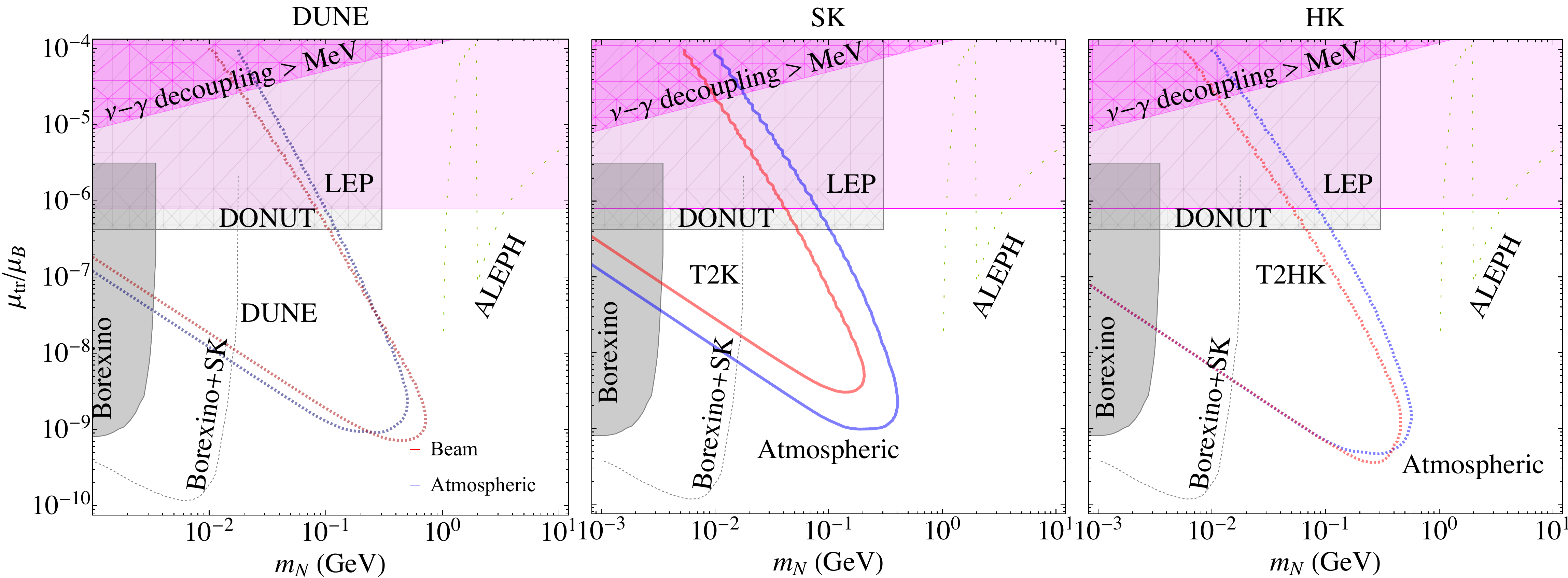}
\caption{Iso-contours indicating the region of parameter space where more than one DB event is expected, for the transition magnetic moment between $\nu_\tau$ and $N$. Results are shown for DUNE (left), Super-K (center), and Hyper-K (right). Same plotting conventions as in Fig.~\ref{fig:HNL}. The shaded areas indicate the regions currently disfavored by Borexino~\cite{Bellini:2014uqa,Brdar:2020quo}, DONUT~\cite{Schwienhorst:2001sj}, and ALEPH~\cite{Decamp:1991uy,Gninenko:2010pr}, and LEP data~\cite{Magill:2018jla}, and Borexino+SK~\cite{Plestid:2020vqf}. The shaded purple area indicates the region leading to delayed neutrino-photon decoupling in the early universe~\cite{Coloma:2017ppo}, and supernovae can constrain smaller couplings~\cite{Magill:2018jla}.}
\label{fig:dipoletau}
\end{center}
\end{figure}

A similar estimate can be performed in the dipole interaction case. From Fig.~\ref{fig:cx} we see that for a HNL with mass $m_{N} = 0.6$ GeV the cross section with $\mu = 10^{-8}~\mu_{B}$ is comparable to the SM NC cross section. Thus using $N_{\nu_{\tau}}^{NC}(E_{\nu} \gtrsim m_{N}) \simeq 4\times 10^{3}$ with Eq.~\eqref{eq:est} roughly suggests that $\mu \gtrsim 10^{-10}~\mu_{B}$ may be within reach for a HNL of mass $m_{N} = 0.6$ GeV. However in this case one can see from the decay length requirement that a value closer to $\mu \gtrsim 10^{-9}~\mu_{B}$ is necessary to decay within the detector volume while still being moderately relativistic, see Eq.~\eqref{eq:lengthdipole}. This roughly agrees with the results from our numerical calculation for the sensitivity using the DUNE nominal beam, shown in Fig.~\ref{fig:dipoletau}. While we have discussed the DUNE case as an example, the same reasoning can be applied to the other cases as well. 

%%%%%%%%%%%%%%%%%%%%%%%%%%%
%%%%%%%%%%%%%%%%%%%%%%%%%%%

Notice that the dominance of the sensitivity for SK atmospheric and the DUNE beam that was already apparent in Fig.~\ref{fig:events} is confirmed in Figs.~\ref{fig:HNL} and \ref{fig:dipoletau}. Moreover, DUNE's dominance at the level of $\nu_{\tau}$ NC events occurs for neutrino energies between roughly $\sim 2-20$ GeV. This explains why the DUNE beam sensitivity to HNLs can exceed the atmospheric Super-K sensitivity at larger HNLs masses, as seen in Figs.~\ref{fig:HNL} and \ref{fig:dipoletau}.

\begin{figure*}[t!]
\begin{center}
\includegraphics[width=.45\columnwidth]{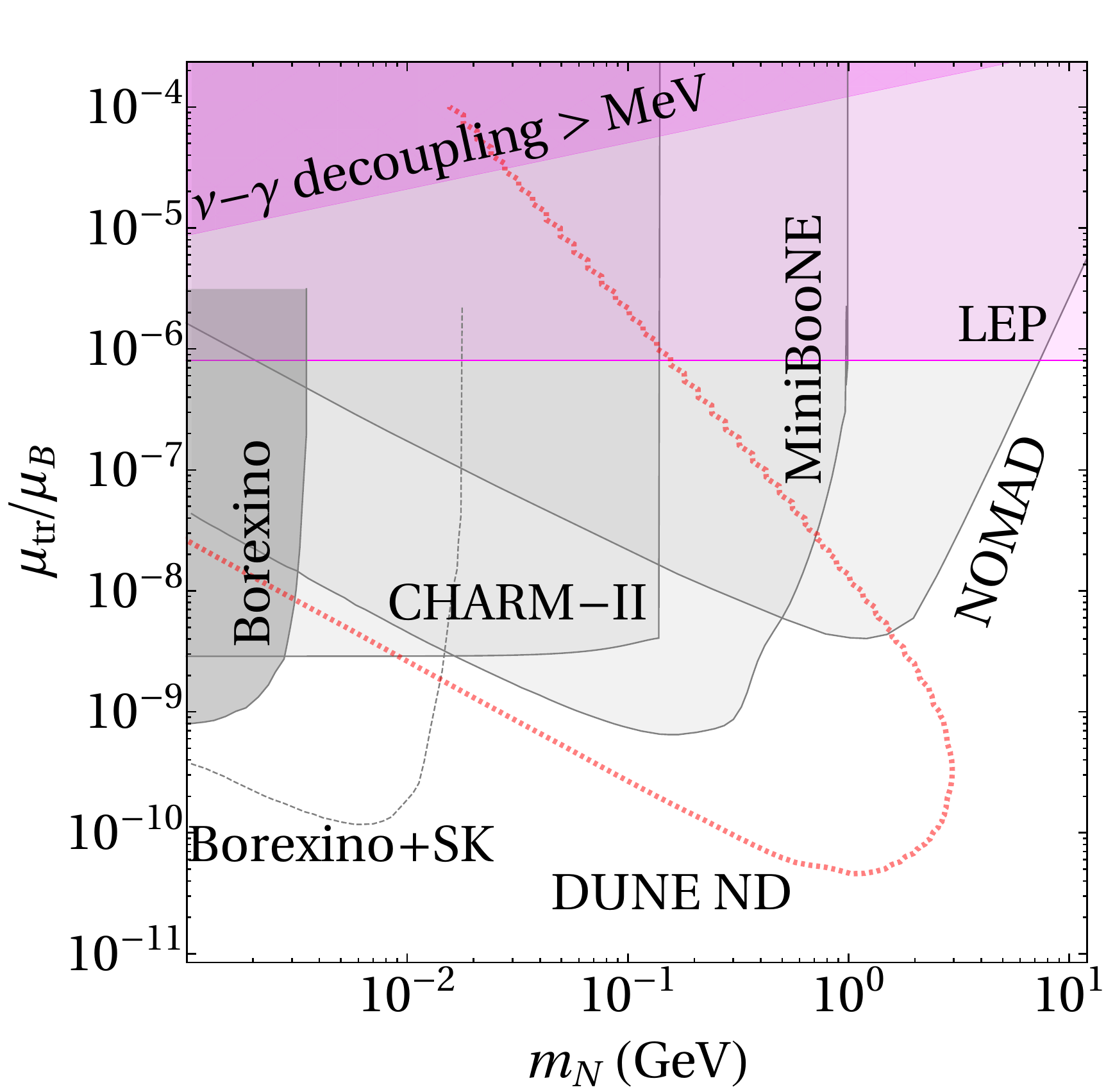}
\caption{Expected sensitivity to the transition magnetic moment $\nu_{\mu}-N$ from DBs signals in the DUNE LAr near detector. The CHARM-II and NOMAD bounds are reproduced from Ref.~\cite{Coloma:2017ppo}. The Borexino+SK bounds are taken from Ref.~\cite{Plestid:2020vqf}}
\label{fig:dipolemu}
\end{center}
\end{figure*}

Lastly, in Fig.~\ref{fig:dipolemu} we show that DUNE's LAr near detector can also be sensitive to DB events. In this case, the large $\nu_{\mu}$ flux at the near detector can allow for $\nu_{\mu}-N$ transition moments to be very well probed, and in fact push into new parameter space which is currently unconstrained. We note that MiniBooNE~\cite{Magill:2018jla} also constrains a small window of parameter space between the CHARM-II and NOMAD bounds in Fig.~\ref{fig:dipolemu}. Based on the sensitivity we find in Fig.~\ref{fig:dipolemu}, we note that the DUNE near detector appears capable of testing a dipole interpretation of the MiniBooNE excess~\cite{Vergani:2021tgc}.

%%%%%%%%%%%%%%%%%%%%%%
%%%%%%%%%%%%%%%%%%%%%%

%%%%%%%%%%%%%%%%%%
\section{Conclusions}
\label{sec:conclusions}
%%%%%%%%%%%%%%%%%%

In this work we have studied the potential of present and future neutrino oscillation experiments, namely, SK, HK and DUNE, to observe double-bang events at low energies, which are characteristic signals of HNL production and decay within the detector. We have considered both atmospheric neutrinos and neutrino beams from meson decays as possible sources for the events, and computed the number of events that would be obtained for the signal in two different scenarios: HNL that mixes with $\nu_\tau$, and a dipole portal involving a transition magnetic moment between the light neutrinos and the HNL. Our results include a detailed calculation of the QE cross section for both scenarios within the spectral function formalism, and a proper treatment of the kinematic effects due to the nonzero HNL mass. 

We have shown that an analysis of current Super-K data may be able to set new and competitive bounds, especially in the HNL mass range 0.3-1 GeV for the mixing scenario. We have also found that DUNE will be able to extend this bound by roughly half an order of magnitude using the DUNE beam flux. This is contrast with the atmospheric neutrinos which do not extend the bounds in the allowed mass range. We have also examined the bounds on active-heavy transition magnetic moments and find that Super-K may be able to considerably improve over the DONUT bounds on tau-flavored transition moments. Similar results are found for the next generation of experiments using neutrinos from the atmosphere and the beam. The large flux of $\nu_{\mu}$ expected in the DUNE near detector allows us to explore a possible transition magnetic moment for muon neutrinos as well. For $1.1\times 10^{22}$~PoT (corresponding to 10 years of data), we find that present bounds can be improved by 2 orders of magnitude, for heavy neutrino masses between 10~MeV and $\sim 1$~GeV.

Our analysis may be improved on several fronts. First, we are only including QE and DIS events in our signal rates, and the inclusion of resonant scattering processes will only improve the sensitivity further. Second, a detailed study of the background levels (such as the punch-through of neutral hadrons, for high-energy events, or the impact of photon conversion to electron/positron pairs) should be performed by the experimental collaborations. In this regard, note that a detailed analysis of the backgrounds for LAr TPC detectors may reveal additional handles which could be used to reduce the minimum separation between the two events while keeping a good signal-to-background ratio, thus enhancing the sensitivity with respect to our results. Finally, a third avenue could be to go beyond the QE and DIS contributions considered here for the magnetic moment scenario and also include the coherent contributions to $\nu \rightarrow N$ up-scattering, where the interaction takes place with the whole nucleus. Although this would considerably enhance the signal rate~\cite{Magill:2018jla,Plestid:2020vqf,Brdar:2020quo}, it may suffer from a larger background as well since the first bang would go unobserved at typical neutrino detectors. A detailed calculation of the signal and background rates needs to be made in order to determine the benefits of coherent scattering for enhancing the sensitivity to HNLs in neutrino experiments, something which is beyond the scope of the present work.

\textbf{Note added:} After the completion of this work the preprint of Ref.~\cite{Schwetz:2020xra} was posted online, where a similar analysis was performed for the DUNE experiment. Taking into account the different cross section calculation used in Ref.~\cite{Schwetz:2020xra}, we find a reasonable agreement with our results.

\vspace{1.5cm}

\textit{Acknowledgements.} 
We warmly thank Justo Mart\'in-Albo for helpful discussions on the separation of events in LAr TPC detectors. The work of IMS is supported by the U.S. Department of Energy under the award number
de-sc0020250. This project has received funding/support from the European Union’s Horizon 2020 research and innovation program under the Marie Skłodowska-Curie grant agreement No 860881-HIDDeN, and  from the Spanish Agencia Estatal de Investigaci\'on through the grant ``IFT Centro de Excelencia Severo Ochoa SEV-2016-0597''. The work of PC is supported by the Spanish MICINN through the ``Ram\'on y Cajal'' program under grant RYC2018-024240-I. TRIUMF receives federal funding via a contribution agreement with the National Research Council of Canada. This manuscript has been authored by Fermi Research Alliance, LLC under Contract No. DE-AC02-07CH11359 with the U.S. Department of Energy, Office of Science, Office of High Energy Physics.

\bibliographystyle{jhep}
\bibliography{nu}

\end{document}